# A Statistical Significance Simulation Study for the General Scientist

**Corresponding Author:** Jacob Levman, PhD

**Author Affiliation:** Sunnybrook Research Institute, University of Toronto, Toronto, ON, Canada. 2075 Bayview Avenue, Room S605, Toronto, ON, Canada, M4N 3M5, 416-480-6100 x3390, fax: 416-480-5714
jacob.levman@sri.utoronto.ca

**Abstract**

When a scientist performs an experiment they normally acquire a set of measurements and are expected to demonstrate that their results are "statistically significant" thus confirming whatever hypothesis they are testing. The main method for establishing statistical significance involves demonstrating that there is a low probability that the observed experimental results were the product of random chance. This is typically defined as $p < 0.05$, which indicates there is less than a 5% chance that the observed results occurred randomly. This research study visually demonstrates that the commonly used definition for "statistical significance" can erroneously imply a significant finding. This is demonstrated by generating random Gaussian noise data and analyzing that data using statistical testing based on the established two-sample t-test. This study demonstrates that insignificant yet "statistically significant" findings are possible at moderately large sample sizes which are very common in many fields of modern science.

**Introduction**

Establishing statistical significance is extremely common among scientists. This involves demonstrating that there is a low probability that a scientist's observed measurements were the result of random chance. This is typically defined as (p<0.05) which indicates that there is less than a 5% chance that the observed differences were the result of randomness. Statistical significance can be established using a wide variety of statistical tests which compare a scientist's measurements with randomly generated distributions to determine a p-value (from which statistical significance is established). It is known that as the number of samples increases, the amount of difference needed between our two distributions to obtain statistical significance (p<.05) gets smaller. The main focus of this research paper is to present data which demonstrates that as the number of samples becomes large, the amount of separation between our two groups needed to obtain 'statistical significance' becomes negligible. This effect indicates that scientists have a potentially extremely low threshold for obtaining statistical significance. In its most extreme form, a "statistically significant" effect is in fact qualitatively insignificant.

In this study we have elected to perform statistical testing using the widely accepted and established two-sample t-test [1]. It should be noted that the t-test was developed in a beer factory in 1908 over one-hundred years ago by a scientist writing under a false name (Student). This was long before the advent of computers, thus long before a scientist had the ability to perform statistical testing on groups of data with large numbers of samples. In fact, the original introduction of the t-test [1] provided look-up tables to assist in statistical computations that allow the researcher to perform analyses on data groups with up to only 10 samples. In those days it was unreasonable for someone to manually compute p-values on hundreds or thousands of samples. In the present research environment a journal paper reviewer is likely to require many more than 10 samples from a typical scientist's experiment, thus inadvertently lowering the bar for obtaining the desired "statistical significance". After reading this study it should be clear that the t-test was developed for another era and alternative techniques would benefit the modern scientist. Or as Bill Rozeboom wrote in 1960 "The statistical folkways of a more primitive past continue to dominate the local scene" [2]. Rozeboom was writing about problems with statistical testing over 50 years after the t-test was first created. It is now 50 years after Rozeboom wrote this commentary and his words are still as relevant as ever. There have been many critiques of how statistical significance testing and null hypothesis testing is used [2-20], yet despite the many shortcomings highlighted, performing hypothesis testing based on a p-value threshold ($p<.05$) is still one of the most common statistical techniques used by modern scientists.

Standard thought has it that if we increase our number of samples then the computed statistical p-value becomes more and more reliable. In fact, as we add more and more samples, the amount of separation needed between our groups to achieve statistical significance gets smaller. This is because the p-value computations are based on random data. Once the number of samples becomes very large, the amount of overlap observed between large randomly generated distributions will always be large, leading to very little separation required between the two distributions to achieve a p-value below 0.05. Or put another way, we have a threshold for statistical significance that is so low that (as long as we have an adequate number of samples) all we need is to have two noisy signals that are ever so marginally dissimilar in order to achieve "statistical significance".

This low threshold for achieving statistical significance has the potential to greatly affect a scientist's approach to their experiments. As scientists, our career prospects (and thus our prestige and personal finances) are heavily dependent on our accumulation of peer-reviewed journal papers. This personal motivation biases us towards getting our research accepted for publication. Since it is extremely common for a journal paper reviewer to require that our experimental results be tested for statistical significance, we are generally biased towards finding statistical significance in our experiments in order to accumulate journal publications and to succeed in our careers. The word 'significant' is qualitative and subjective. Whether something is 'significant' is in the eye of the beholder. When we add the word 'statistics', we add a strong quantitative word to the very qualitative word 'significant'. This lends an appearance of credibility and certainty to any experiment that achieves a p-value below 0.05, simply because this is the widely accepted threshold for achieving 'statistical significance'.

Since statistical significance is based on random distributions, performing hypothesis testing on the p-value calculation (p<.05) is like asking the question: did our experiment do better than 95% of randomness? But since the vast majority of scientists are likely to have constructed their experiments in a somewhat logical manner, they are generally liable to do at least a little better than random chance. Thus scientists are highly likely to find statistical significance in their experiments, especially if they perform their experiments with many samples. This study is designed to visually illustrate that achieving statistical significance (p<.05) at moderately large sample sizes requires only marginally significant (or possibly even insignificant) experimental data.

**Methods**

P-values are computed from lookup tables which are created from randomly generated distributions of data. This research study's methods are designed to visually illustrate how much separation is required between two groups of numbers in order to achieve the standard definition of statistical significance (p<.05) at a variety of sample sizes. This is accomplished by generating large amounts of normal (Gaussian) random distributions. We have elected to perform our analysis on two-sample tests where two groups of numbers are compared with each other in order to determine if they are statistically significantly different from each other. This is one of the most pervasive types of statistical testing as it is extremely common for a scientist to compare two groups of numbers (for example an experimental group and a control group). For this study 1000 pairs of random distributions were created at each example sample size. Of all the randomly generated cases, the pair that exhibit the highest p-value below 0.05 is selected for presentation as a visual example of how much separation is needed between two groups of data in order to achieve 'statistical significance' at the given sample size. Random distributions were generated across a wide variety of group sample sizes where the image's dimensions are expressed as a factor of 2 (4, 16, 64, 256, 1024, 4096, 16384, 65536 and 262144 samples in the example distributions presented). The variance in these noise pairs demonstrates how the amount of separation between two barely statistically significantly different groups changes as the number of samples is varied.

All statistical significance testing was performed using one of the most common statistical tests available, the two-sample t-test. This was selected so that our statistical testing method matches the type of distributions being randomly generated (Gaussian noise / normal distributions). In addition, for each sample size setting, the number of randomly created distributions that have a p-value below 0.05 are enumerated. All random normal (Gaussian) distributions were created using the mathematical and statistical package Matlab (Mathworks, Natick, MA, USA). Statistical testing was performed with the established two-sample t-test provided in Matlab.

**Results**

Pairs of randomly generated distributions with p-values just below 0.05 are included as the main results of this study. Figure 1 demonstrates randomly generated statistically significantly different normal distributions with 4 samples (2x2, top row), 16 samples (4x4, middle row) and 64 samples (8x8, bottom row). Figure 2 demonstrates randomly generated statistically significantly different normal distributions with 256 samples (16x16, top row), 1024 samples (32x32, middle row) and 4096 samples (64x64, bottom row). Figure 3 demonstrates randomly generated statistically significantly different normal distributions with 16384 samples (128x128, top row), 65536 samples (256x256, middle row) and 262144 (512x512, bottom row). Table 1 presents the p-values of each of the pairs selected for viewing in figures 1, 2 and 3 as computed by Matlab's two-sample t-test. Table 1 also presents the total number of randomly generated distributions which achieved a statistically significant difference as the term is typically defined (p<.05), using the popular and well established two-sample t-test. Since the experiment involves creating 1000 randomized distributions we expect to find 50 (5%) of those samples being statistically significant (p<.05). The results from each trial were confirmed to be close to 50 samples achieving statistical significance out of each 1000 randomly created cases.

When examining each noise image pair, a scientist can interpret the two visual image distributions as being very close to the threshold for obtaining statistical significance at the given number of samples. Note that the difference between two statistically significantly different distributions gets smaller as the number of samples increases.

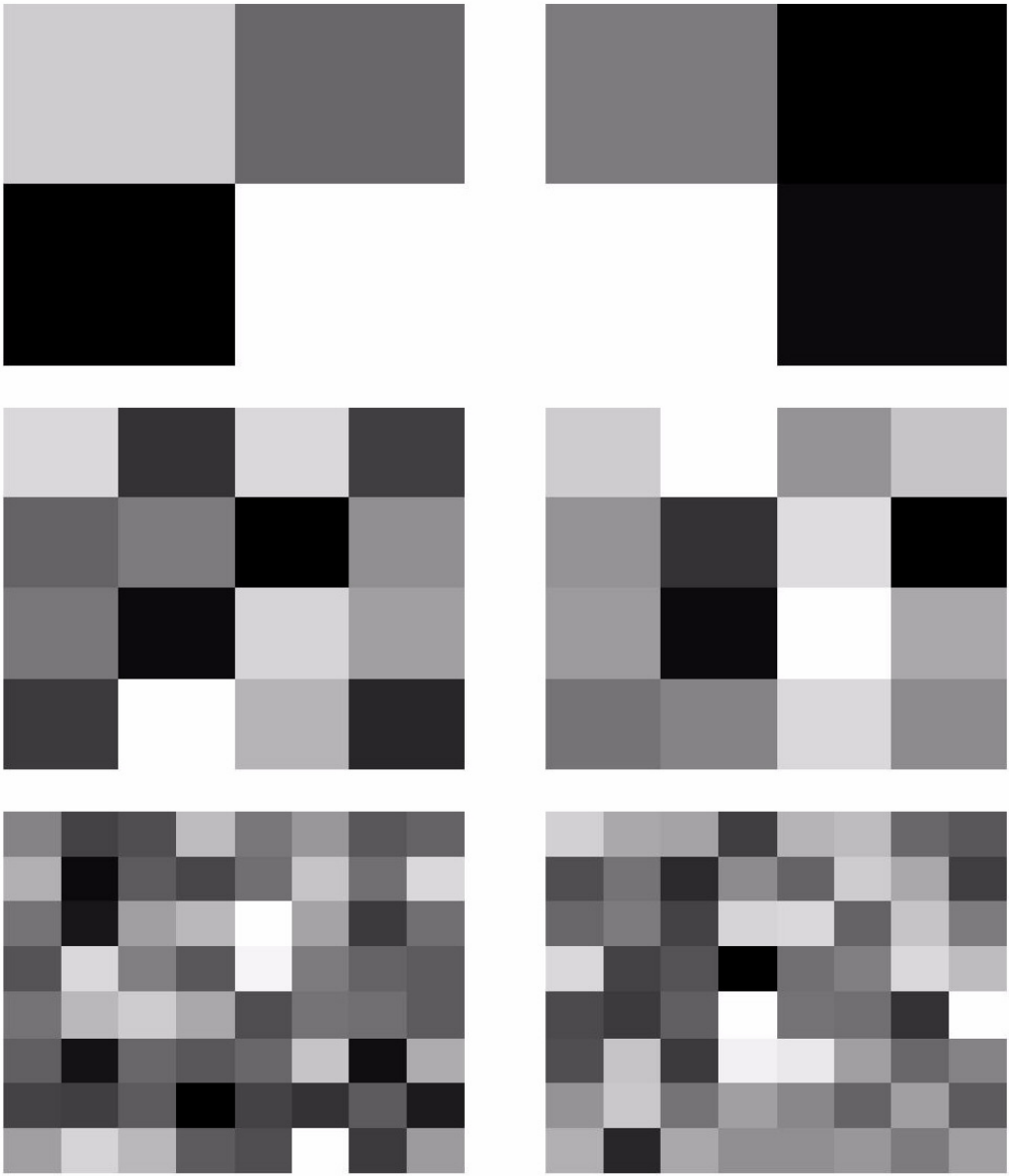

Figure 1: Randomly generated pairs of statistically significantly different (p<.05) distributions with 4 samples (top row), 16 samples (middle row) and 64 samples (bottom row).

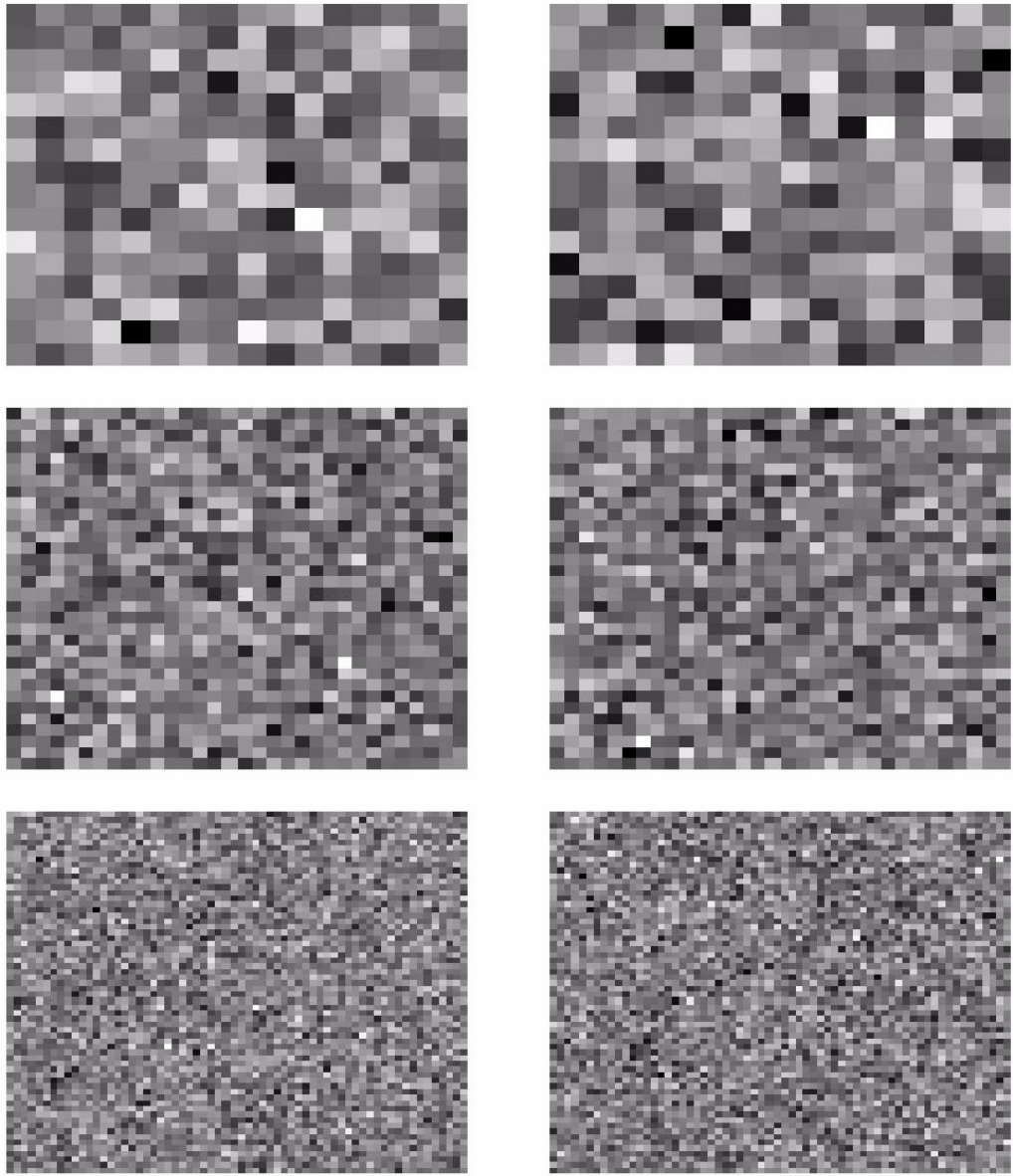

Figure 2: Randomly generated pairs of statistically significantly different (p<.05) distributions with 256 samples (top row), 1024 samples (middle row) and 4096 samples (bottom row).

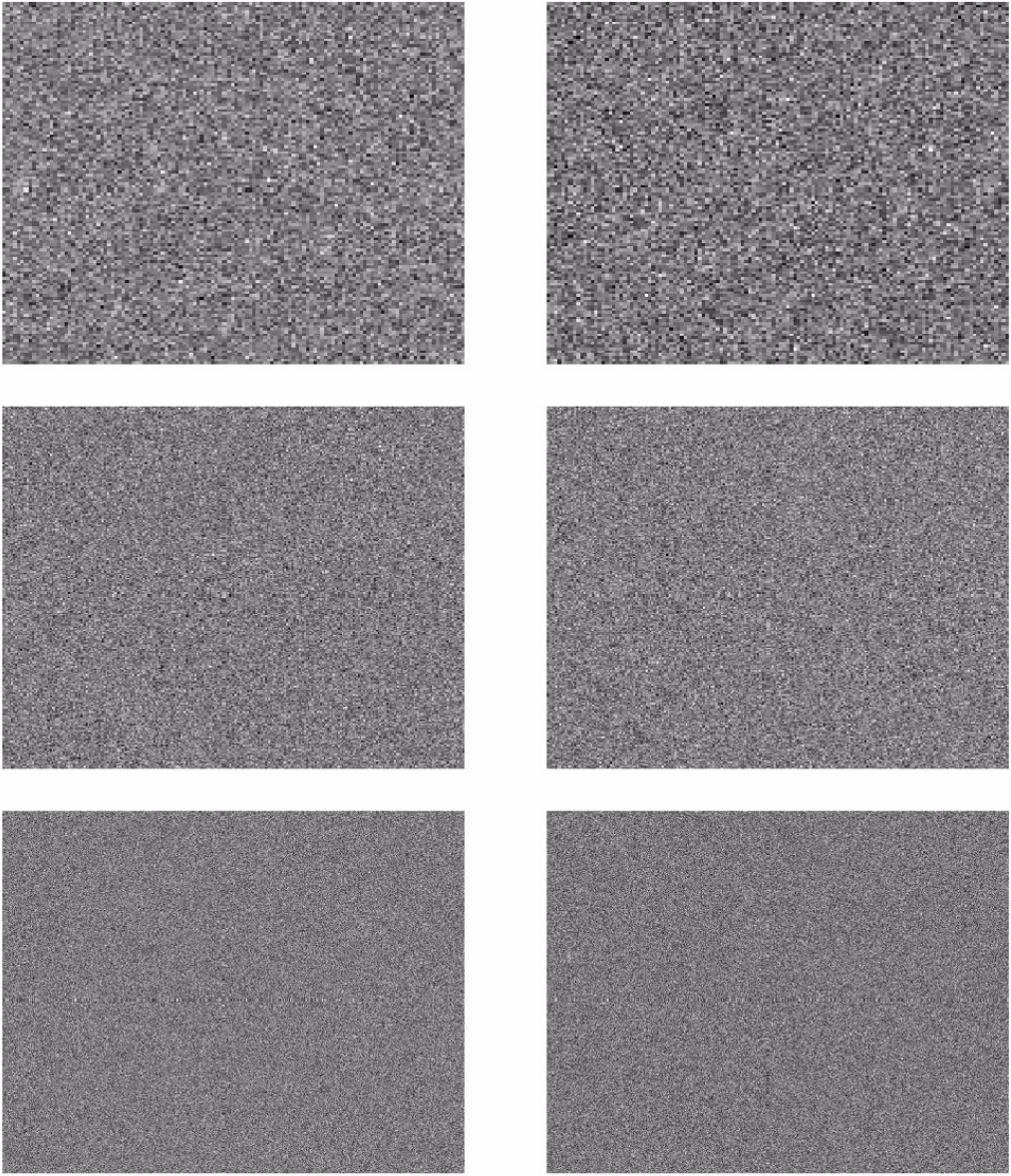

Figure 3: Randomly generated pairs of statistically significantly different (p<.05) distributions with 16384 samples (top row), 65536 samples (middle row) and 262144 samples (bottom row).

Table 1: P-values and Associated Data for the Randomly Generated Distributions of Figures 1, 2 and 3

| Random Distribution Size | P-value of Pair Presented in Figures Above | Number of Random Cases with p < 0.05 |
|---|---|---|
| 2x2=4 | 0.0499 | 51/1000 |
| 4x4=16 | 0.0459 | 44/1000 |
| 8x8=64 | 0.0488 | 48/1000 |
| 16x16=256 | 0.0493 | 49/1000 |
| 32x32=1024 | 0.0498 | 54/1000 |
| 64x64=4096 | 0.0499 | 56/1000 |
| 128x128=16384 | 0.0483 | 46/1000 |
| 256x256=65536 | 0.0485 | 42/1000 |
| 512x512=262144 | 0.0496 | 42/1000 |

**Discussion**

It can be seen from the results that at a p-value just below 0.05, the two randomly generated groups of 4 samples each are substantially different from each other as the image on the right is clearly darker overall than the image on the left (see figure 1 top line). At 16 and 64 samples, it is clear that the random image on the left is darker than the one on the right although it is clear that the 64 sample images are substantially more similar to each other than the images with 16 or 4 samples. All of the statistically significant pairs presented in Figure 1 appear qualitatively significantly different from each other. Once the size of the images has been increased to just 256 samples it becomes challenging to see a significant difference between the two distributions, even though the results displayed are statistically significant (p = 0.0493) as the term is traditionally used

(see figure 2 top line). When comparing two distributions with more than 256 samples, the distributions appear qualitatively insignificantly different from each other despite having obtained "statistical significance" (see figures 2 and 3).

Data was also included demonstrating that approximately 5% of the randomly generated samples created for this experiment achieve statistical significance (as the term is typically defined p <.05). This is presented in the final column of Table 1. This information is simply provided to demonstrate that the experiment is matching expectations – that about 5% or about 50 out of 1000 randomly created distributions have a p-value below 0.05.

P-value computations from single sample statistical tests are intuitive: a new single sample can be compared against the pre-existing group and the 0.05 p-value threshold causes only those samples that fall on the outskirts of the distribution to be considered statistically significantly different. The same does not hold once we move to two-sample statistical testing. If we generate two random groups of data with each group containing many samples, then it is inevitable that the two groups will overlap each other substantially. Even in the 5% of cases where the two large random distributions are most dissimilar, we will still find highly overlapping distributions as demonstrated in this paper's results (figures 2 and 3). This has the effect of setting the bar for finding statistical significance in our experiments extremely low (especially when the number of samples is large) and may have led researchers to conclude a significant effect from their experimental results when in fact the effect observed is much smaller or possibly even non-existent. Achieving statistical significance (p<.05) merely demonstrates that the experimental results outperformed 95% of the randomly generated noise from which the p-value is computed. Ascribing 'significance' to any experiment is a subjective task which should be evaluated by whomever is interested in examining the experiment, not by a single number.

This study's findings are potentially of broad interest to scientists in general. Figure 2 (top line) demonstrates that achieving statistical significance (p<.05) on groups with only 256 samples only confirms the existence of an extremely marginal effect. Scientific studies based on at least a couple hundred samples in each group are extremely common in the literature. Establishing statistical significance is typically a prerequisite for publication of a scientific study. Scientists who find statistical significance in experiments containing thousands of samples haven't actually demonstrated that their findings are significant at all (unless they've included well separated confidence intervals). It is doubtful that anyone would qualitatively describe the pairs of results presented in figure 3 as significantly different from each other even though they meet the normal criteria for statistical significance (p<.05).

Establishing statistical significance with a p-value provides us with an answer to the question "did we beat 95% of randomness?" But randomness is a very low bar to set for ourselves, thus ensuring that scientists who work with reasonably large sample sizes will be able to go on finding statistically significant (p<.05) results (almost) wherever they look for them.

**Acknowledgments**

This work was supported by the *Canadian Breast Cancer Foundation*.